\documentclass[aps,pra,showpacs,superscriptaddress,preprint]{revtex4}

\usepackage{graphicx}

\catcode`ð=\active

\defð{\u{g}}

\catcode`Ð=\active

\defÐ{\u{G}}

\catcode`Ý=\active

\defÝ{\. I}

\catcode`ö=\active

\defö{\"{o}}

\catcode`Ö=\active

\defÖ{\"O}

\catcode`ü=\active

\defü{\"{u}}

\catcode`Ü=\active

\defÜ{\"{U}}

\catcode`Þ=\active

\defÞ{\c{S}}

\catcode`þ=\active

\defþ{\c{s}}

\catcode`ý=\active

\defý{{\i}}

\catcode`ç=\active

\defç{\d{c}}

\catcode`Ç=\active

\defÇ{\d{C}}

\begin{document}
\title{Analytical Solutions of Klein-Gordon Equation with Position-Dependent
Mass for $q$-Parameter P\"{o}schl-Teller potential }
\author{\small Altuð Arda}
\email[E-mail: ]{arda@hacettepe.edu.tr}\affiliation{Department of
Physics Education, Hacettepe University, 06800, Ankara,Turkey}
\author{\small Ramazan Sever}
\email[E-mail: ]{sever@metu.edu.tr}\affiliation{Department of
Physics, Middle East Technical  University, 06800, Ankara,Turkey}
\author{\small Cevdet Tezcan}
\email[E-mail: ]{ctezcan@baskent.edu.tr}\affiliation{Faculty of
Engineering, Baþkent University, Baglýca Campus, Ankara,Turkey}

\date{\today}

\begin{abstract}

The energy eigenvalues and the corresponding eigenfunctions of the
one-dimensional Klein-Gordon equation with $q$-parameter
P\"{o}schl-Teller potential are analytically obtained within the
position-dependent mass formalism. The parametric generalization
of the Nikiforov-Uvarov method is used in the calculations by choosing a mass distribution.
\end{abstract}

\maketitle

\newpage

\section{Introduction}

In last few years, the formalism constructed on varying of the
mass with coordinates has been received much attentions because of
finding many applications in different areas. The
position-dependent mass formalism has been used to describe the
electronic properties of semiconductors and quantum dots [1-5]. It
also gives interesting results in the study of $^{3}$He clusters
[6], and quantum liquids [7].

The position-dependent mass formalism brings some basic problems
in the investigation of physical systems, such as the possible
discontinuities of choosen mass functions, the invariance of
theory under the Galileo transformations, and ordering-ambiguity
between momentum, and mass operators in kinetic energy term [8,
9].

Recently, the solutions of non-relativistic wave equation with
constat mass have been extended to the position-dependent mass
(PDM) case [10-25]. In Ref. [26], a general formalism giving both
of energy spectra and also wave functions were found in
non-relativistic problems by mapping between reference and target
systems by using point canonical transformation. The
Schr\"{o}dinger equation with position-dependent mass is studied
by using parameter algebra, and in curved spaces has been
investigated in the Coulomb example [27]. In Ref. [28], the
shape-invariance technique has been applied to solvable
Hamiltonians. Bagchi, and co-workers have generated a formalism to
obtain $PT$-symmetric Hamiltonians in the PDM case [29]. In Re.
[30] has been made a generalization of the point canonical
transformation to solve the Schr\"{o}dinger equation in
$d$-dimension. The authors have also investigated the
$\eta$-pseudo-hermicity of non-Hermitian Hamiltonians with
position-dependent mass [31]. The investigation of a non-Hermitian
Hamiltonian for an oscillator defined by Swanson's model has been
made in the context of N-fold supersymmetry [32].

Another area received much attentions is the extension of the
non-relativistic solutions to the relativistic solutions in the
view of position-dependent mass formalism. The energy spectra and
corresponding wave functions of Klein-Gordon and Dirac equations
have been found for different potentials by using different
methods, such as complexified Lorentz scalar interactions [33],
Coulomb potential [34], hyperbolic potentials [35], non-Hermitian
complexified potentials [36, 37], $PT$-symmetric trigonometric
potential [38], $PT$-symmetric harmonic oscillator-like,
$PT$-symmetric inversely linear plus linear, and $PT$-symmetric
kink-like potentials [39], inversely linear plus linear potential,
and Scarf II potential [40].

Quantum deformation [41] has received much attentions because of
its relation with applications in nuclei [42-45],
statistical-quantum theory, string/brane theory and conformal
field theory [46-49]. Recently, some authors have been introduced
some potentials in terms of hyperbolic functions in the view of
$q$-deformation [50]. One of the potential from this family can be
written as

\begin{eqnarray}
V(x;q)=-\frac{V_0}{cosh^2_{q} (\alpha x)}\,,
\end{eqnarray}
which is the $q$-parameter form of the usual P\"{o}schl-Teller
potential [51]. Here, $q$ is the deformation parameter and used to
denote a mapping from a $c$-number $N$ to a $q$-number $[N]_p$ by
relation

\begin{eqnarray}
[N]_{p}=\frac{(e^{\kappa N})^{p}-(e^{\kappa
N})^{-p}}{(e^{\kappa})^{p}-(e^{\kappa})^{-p}}\,\rightarrow N\,,
\end{eqnarray}
in the $q \rightarrow 1$ limit, $\kappa$ is a real parameter, and
$e^{\kappa}=q$ in the above equation [41]. The $q$-parameter
hypergeometric functions in Eq. (1) are defined

\begin{eqnarray}
sinh_q (z)=\,\frac{1}{2}(e^z-qe^{-z})\,\,\,;\,\,\,cosh_q
(z)=\,\frac{1}{2}(e^z+qe^{-z})\,,\\
tanh_q (z)=\frac{e^z-qe^{-z}}{e^z+qe^{-z}}\,\,\,;\,\,\,sech_q
(z)=\frac{2}{e^z+qe^{-z}}\,.
\end{eqnarray}

In the present work, we intend to solve analytically the
one-dimensional effective Klein-Gordon equation for $q$-parameter
P\"{o}schl-Teller potential to investigate the effect of the PDM
to the energy spectra and corresponding wave functions. We choose
an exponentially mass distribution function which makes it
possible to analytically solve the Klein-Gordon equation. We use
the parametric generalization of the Nikiforov-Uvarov (NU) method
[35] to find the energy eigenvalues, and corresponding wave
functions [53]. The NU method describes the way to solve a
Schr\"{o}dinger-like equation by turning it into a hypergeometric
type equation [52].

The organization of this work is as follows. In Section II, we
solve the one-dimensional Klein-Gordon equation for $q$-parameter
P\"{o}schl-Teller potential, and give the energy spectrum, and the
corresponding wave functions in the case of position-dependent
mass. We summarize our conclusion in Section III.

\section{Bound-States of Klein-Gordon Equation}

The one-dimensional Klein-Gordon equation for a particle with mass
$m$ subject to scalar, $V_s (x)$, and vector, $V_v (x)$,
potentials reads ($\hbar=c=1$)

\begin{eqnarray}
\frac{d^2\psi(x)}{dx^2}\,+\left[\left[E-V_{v}(x)\right]^2-
\left[m-V_{s}(x)\right]^2\right]\psi(x)=0\,,
\end{eqnarray}
where $E$ is the relativistic energy of particle.

We prefer to use the mass function equal to the vector part of the
potential as

\begin{eqnarray}
m(x)=m_0+4V_0\,\frac{e^{-2\alpha x}}{(1+qe^{-2\alpha x})^2}\,,
\end{eqnarray}
to obtain an exactly solvable Schr\"{o}dinger-like equation from
Eq. (5) in the absence of scalar potential. The mass function
should also be a physically distribution, so we restrict ourself
in the range $0 \leq x \leq \infty$\,, which gives the following
finite mass values

\begin{displaymath}
m(x)=\left\{ \begin{array} {ll} m_{0}+2V_{0}\,\,(\textrm{for}\,\,q
\rightarrow 1), & x \rightarrow 0\,,\\ m_{0}, & x \rightarrow
\infty\,.
\end{array} \right.
\end{displaymath}

Indeed, this distribution corresponds to shifted scalar potential
function in the problem. Substituting Eq. (1), and Eq. (6) into
Eq. (5), we get

\begin{eqnarray}
\frac{d^2\psi(x)}{dx^2}+\Big\{\,(E-m_0+8V_{0}\,\frac{e^{-2\alpha
x}}{(1+qe^{-2\alpha x})^2})(E+m_0)\Big\}\psi(x)=0\,,
\end{eqnarray}
By using the new variable $s=e^{-2\alpha x} (0<s<1)$, we have

\begin{eqnarray}
\frac{d^2\psi(s)}{ds^2}&+&\frac{1+qs}{s(1+qs)}\frac{d\psi(s)}{ds}+\frac{1}{[s(1+qs)]^2}\nonumber\\
&\times&\Big\{\eta^2q^2(E^2-m^2_0)s^2+[2\eta^2q(E^2-m^2_0)+8\eta^2V_{0}(E-m_0)]s\nonumber\\
&+&\eta^2(E^2-m^2_0)\Big\}\psi(s)=0\,,
\end{eqnarray}
where $\eta^2=1/4\alpha^2$\,. Following Ref. [36], we obtain the
parameter set as

\begin{eqnarray}
\begin{array}{ll}
\alpha_1=1\,, & -\xi_1=\eta^2q^2(E^2-m^2_0) \\
\alpha_2=-q\,, &
\xi_2=2\eta^2q(E^2-m^2_0)+8\eta^2V_{0}(E-m_0) \\
\alpha_3=-q\,, &
-\xi_3=\eta^2(E^2-m^2_0) \\ \alpha_4=0\,, & \alpha_5=\frac{q}{2} \\
\alpha_6=\xi_1+\frac{q^2}{4}\,, & \alpha_7=-\xi_2 \\
\alpha_8=\xi_3\,, & \alpha_9=\xi_1+q\xi_2+q^2\xi_3+\frac{1}{4}\,q^2 \\
\alpha_{10}=1+2\sqrt{\xi_3}\,, & \alpha_{11}=-2q+2(\,\sqrt{\xi_1+q\xi_2+q^2\xi_3+\frac{1}{4}\,q^2\,}-q\sqrt{\xi_3}\,) \\
\alpha_{12}=\sqrt{\xi_3}\,, &
\alpha_{13}=\frac{q}{2}-(\,\sqrt{\xi_1+q\xi_2+q^2\xi_3+\frac{1}{4}\,q^2\,}-q\sqrt{\xi_3}\,)
\end{array}
\end{eqnarray}
and we deduce the parameters required for the method [36]

\begin{eqnarray}
\pi(s)&=&\frac{1}{2}\,q \pm
\bigg\{\bigg[\frac{1}{4}\,q^2-2q\sqrt{\eta^2(m^2_0-E^2)[8V_0\eta^2q(E-m_0)+\frac{1}{4}\,q^2]\,}\bigg]s^2\nonumber\\
&-&\bigg[2\eta^2q(E^2-m^2_0)-2\sqrt{\eta^2(m^2_0-E^2)[8V_0\eta^2q(E-m_0)+\frac{1}{4}\,q^2]\,}\bigg]s
\nonumber\\&-&\eta^2(E^2-m^2_0)\bigg\}^{1/2}\,,
\end{eqnarray}
and
\begin{eqnarray}
k&=&-2\sqrt{\eta^2(m^2_0-E^2)[8V_0\eta^2q(E-m_0)+\frac{1}{4}\,q^2]\,}\,,\\
\tau(s)&=&1-2qs-2\bigg\{\bigg[\sqrt{8V_0\eta^2q(E-m_0)+\frac{1}{4}\,q^2\,}-q\sqrt{\eta^2(m^2_0-E^2)\,}\,\bigg]s
\nonumber\\&-&\sqrt{\eta^2(m^2_0-E^2)\,}\bigg\}\,.
\end{eqnarray}
We need to know $\lambda$ and $\lambda_{n}$ [36]

\begin{eqnarray}
\lambda&=&-2\sqrt{\eta^2(m^2_0-E^2)[8V_0\eta^2q(E-m_0)+\frac{1}{4}\,q^2]\,}+\pi'(s)\,,\\
\lambda_n&=&-n\bigg[-2q-2\sqrt{8V_0\eta^2q(E-m_0)+\frac{1}{4}\,q^2\,}+2q\sqrt{\eta^2(m^2_0-E^2)\,}\,\bigg]-qn(n-1)\,,
\end{eqnarray}
to get the eigenvalue equation for the energy spectra of the
$q$-parameter P\"{o}schl-Teller potential as

\begin{eqnarray}
\frac{1}{\alpha}\,\sqrt{m^2_0-E^2\,}+\sqrt{\frac{1}{4}+\frac{2V_0}{q\alpha^2}(E-m_0)\,}=n+\frac{1}{2}\,,
\end{eqnarray}
which gives two independent solutions corresponding to particle
and antiparticle states. This equation gives $E=\pm m_{0}$ for the
constant mass case when the potential vanishes.

The corresponding unnormalized eigenfunctions are obtained in
terms of following functions [36]

\begin{eqnarray}
\rho(s)&=&s^{2\sqrt{\eta^2(m^2_0-E^2)\,}}\,(1+qs)^{-2\sqrt{\frac{8}{q}\,V_0\eta^2(E-m_0)+\frac{1}{4}\,}}\,,\\
y_n(s)&=&P_n^{(\,2\sqrt{\eta^2(m^2_0-E^2)\,}\,,-2\sqrt{\frac{8}{q}\,V_0\eta^2(E-m_0)+\frac{1}{4}\,})}(1+2qs)\,,\\
\end{eqnarray}
and
\begin{eqnarray}
\phi(s)=s^{\sqrt{\eta^2(m^2_0-E^2)\,}}(1+qs)^{-\frac{1}{2}-\sqrt{\frac{8}{q}\,V_0\eta^2(E-m_0)+\frac{1}{4}\,}}\,,
\end{eqnarray}
as

\begin{eqnarray}
\psi_n
(s)&=&a_{n}s^{\sqrt{\eta^2(m^2_0-E^2)\,}}(1+qs)^{-\frac{1}{2}-\sqrt{\frac{8}{q}\,V_0\eta^2(E-m_0)+\frac{1}{4}\,}}
\nonumber\\
&\times&P_n^{(\,2\sqrt{\eta^2(m^2_0-E^2)\,}\,,-2\sqrt{\frac{8}{q}\,V_0\eta^2(E-m_0)+\frac{1}{4}\,})}(1+2qs)\,.
\end{eqnarray}
where $a_{n}$ is a normalization constant. The eigenfunctions are
dependent on the Jacobi polynomials
$P^{(\alpha\,,\beta)}_{n}(x)$\,. The asymptotic behavior of the
wave function could be given in the limit $s \rightarrow 0$ as

\begin{eqnarray}
\psi_n(s) \rightarrow 0\,,
\end{eqnarray}
because
$P_{n}^{(\alpha\,,\beta)}(1)=\frac{\Gamma(\alpha+n+1)}{n!\Gamma(1+\alpha)}$
[54], and we write the wave function for $s \rightarrow 1$ as

\begin{eqnarray}
\psi_n(s) \sim P_{n}^{(\alpha\,,\beta)}(1+2q)\,,
\end{eqnarray}
where a physical solution can be obtained only for $-1<q\leq1$.

\section{Conclusion}
We have solved analytically the Klein-Gordon equation for the
$q$-parameter P\"{o}schl-Teller potential in one-dimension in the
case of position-dependent mass function. We obtain a energy
eigenfunction, which gives particle and antiparticle states, by
using the parametric generalization of the NU-method, and get the
corresponding eigenfunctions in terms of Jacobi polynomials. We
also study the energy spectra for the case where mass is constant
and potential vanishes. Finally, we give the behavior of the
corresponding wave function at zero.

\section{Acknowledgments}

This research was partially supported by the Scientific and
Technical Research Council of Turkey.


\newpage

\begin{thebibliography}{99}

\bibitem{1}  J.~C.~Slater, Phys. Rev. {\bf 76}, 15932 (1949).

\bibitem{2} G.~H.~Wannier, Phys. Rev. {\bf 52}, 191 (1937).

\bibitem{3} J.~M.~Luttinger, and W.~Kohn, Phys. Rev. {\bf 97}, 869
(1955).



\bibitem{4} L.~Serra, and E.~Lipparini, Europhys. Lett. {\bf 40}, 667 (1997).



\bibitem{5} G.~Bastard, Wave Mechanics Applied to Semiconductor Hetrostructures (Les Ulis: Editors de Physique, 1998).



\bibitem{6} M.~Barranco, M.~Pi, S.~M.~Gatica, E.~S.~Hernandez, and J.~Navarro, Phys. Rev. B {\bf 56}, 8997 (1997).



\bibitem{7} F.~A.~de Saavedra, J.~Boronat, A.~Polls, and A.~Fabrocini, Phys. Rev. B {\bf 50}, 4248 (1994).



\bibitem{8} J.~M.~L\'{e}vy-Leblond, Phys. Rev. A {\bf 52}, 1845 (1995).



\bibitem{9} F.~S.~A.~Cavalcante, R.~N.~C.~Filho, H.~R.~Filho, C.~A.~S.~Almeida, and V.~N.~Freire,
Phys. Rev. B {\bf 55}, 1326 (1997).



\bibitem{10} L.~Dekar, L.~Chetouani, and T.~F.~Hamann, J. Math. Phys. A {\bf 39}, 2551
(1998); Phys. Rev. A {\bf 59}, 107 (1999).

\bibitem{11} A.~R.~Plastino, A.~Rigo, M.~Casas, F.~Gracias, and A.~Plastino, Phys. Rev. A {\bf
60}, 4318 (1999).

\bibitem{12} A.~S.~Dutra, and C.~A.~S.~Almeida, Phys. Lett. A
{\bf 275}, 25 (2000).

\bibitem{13} B.~Roy, and P.~Roy, J. Phys. A {\bf 35},
3961 (2002).

\bibitem{14} S.~H.~Dong, and M.~Lozada-Cassou, Phys. Lett A {\bf
337}, 313 (2005).

\bibitem{15} J.~Yu, and S.~H.~Dong, Phys. Lett A {\bf 325},
194 (2004).

\bibitem{16} J.~Yu, S.~H.~Dong, and G.~H.~Sun, Phys. Lett A {\bf
322}, 290 (2004).

\bibitem{17} A.~S.~Dutra, M.~Hott, and C.~A.~S.~Almeida, Europhys. Lett. {\bf 62}, 8 (2003).

\bibitem{18} R.~Koc, M.~Koca, and E.~Korcuk, J. Phys. A {\bf 35}, L527 (2002).

\bibitem{19} R.~Koc, and M.~Koca, J. Phys. A {\bf 36}, 8105 (2003).

\bibitem{20} B.~Gonul, B.~Gonul, D.~Tutcu, and O.~Ozer, Mod. Phys. Lett A {\bf 17}, 2057 (2002).

\bibitem{21} B.~Gonul, O.~Ozer, B.~Gonul, and F.~Uzgun, Mod. Phys. Lett. A
{\bf 17}, 2453 (2002).

\bibitem{22} R~Sever, and C.~Tezcan, Int. J. Mod. Phys. E {\bf 17}, 1327 (2008) [arXiv:quant-ph/0712.0268].

\bibitem{23} C.~Tezcan, and R.~Sever, Int. J. Theor. Phys. {\bf 47}, 1471 (2008).

\bibitem{24} S.~Ikhdair, and R.~Sever, [arXiv:quant-ph/0604095].

\bibitem{25} C.~Tezcan, and R.~Sever, J. Math. Chem.
{\bf 42}, 387 (2007).

\bibitem{26} A.~D.~Alhaidari, Phys. Rev. A {\bf 66}, 0421116 (2002).

\bibitem{27} C.~Quesne, and V.~M.~Tkachuk, J. Phys. A {\bf 37}, 4267 (2004).

\bibitem{28} B.~Bagchi, A.~Banerjee, C.~Quesne, and  V.~M.~Tkachuk, J. Phys. A {\bf 38}, 2929 (2005).

\bibitem{29} B.~Bagchi, C.~Quesne, and R.~Roychoudhury, J. Phys. A
{\bf 39}, L127 (2006).

\bibitem{30} O.~Mustafa, and S.~H.~Mazharimousavi, J. Phys. A {\bf 39},
10537 (2006).

\bibitem{31} O.~Mustafa, and S.~H.~Mazharimousavi, Czech.
J. Phys. {\bf 56}, 967 (2006).

\bibitem{32} B.~Bagchi, and T.~Tanaka, Phys. Lett. A {\bf 372}, 5390 (2008).

\bibitem{33} O.~Mustafa, and S.~H.~Mazharimousavi, Int. J. Theor. Phys. {\bf 47},
1112 (2008).

\bibitem{34} A.~D.~Alhaidari, Phys. Lett. A {\bf 322}, 12 (2004).

\bibitem{35} C.~S.~Jia, P.-Q.~Wang, J.-Y.~Liu, and S.~He, Int. J. Theor. Phys. {\bf 47}, 2513 (2008).



\bibitem{36} C.~S.~Jia, and A.~S.~Dutra, J. Phys. A {\bf 39}, 11877
(2006).

\bibitem{37} O.~Mustafa, and S.~H.~Mazharimousavi, J. Phys. A {\bf 40},
863 (2007).

\bibitem{38} C.~S.~Jia, J.-Y.~Liu, P.-Q.~Wang, C.~S.~Che, Phys. Lett. A
{\bf 369}, 274 (2007).

\bibitem{39} C.~S.~Jia, and A.~S.~Dutra, Ann. Phys. {\bf 323}, 566 (2008).

\bibitem{40} O.~Mustafa, and S.~H.~Mazharimousavi, Int. J. Theor. Phys. {\bf 47}, 446
(2008).


\bibitem{41} M.~Jimbo, Lett. Math. Phys. {\bf 10}, 63 (1985); {\bf 11}, 247 (1986).

\bibitem{42} K.~D.~Sviratcheva, C.~Bahri, A.~I.~Georgieva, and J.~P.~Draayer, Phys. Rev. Lett. {\bf 93}, 152501 (2004).

\bibitem{43} A.~Ballesteros, O.~Civitarese, F.~J.~Herranz, and M.~Reboiro, Phys. Rev. C {\bf 66}, 064317 (2002).

\bibitem{44} A.~Ballesteros, O.~Civitarese, and M.~Reboiro, Phys. Rev. C
{\bf 68}, 044307 (2003).

\bibitem{45} D.~Bonatsos, B.~A.~Kotsos, P.~P.~Raychev, and
P.~A.~Terziev, Phys. Rev. C {\bf 66}, 054306 (2002).

\bibitem{46} A.~Ballesteros, N.~R.~Bruno, and F.~J.~Herranz, Phys.
Lett. B {\bf 574}, 276 (2003).

\bibitem{47} F.~A.~Bais, B.~J.~Schroers, and J.~K.~Slingerland,
Phys. Rev. Lett. {\bf 89}, 181601 (2002).

\bibitem{48}  A.~Algýn, M.~Arýk, and A.~S.~Arýkan, Phys. Rev. E {\bf 65}, 026140 (2002).

\bibitem{49} J.~Zhang, Phys. Lett. B {\bf 477}, 361 (2000).

\bibitem{50} A.~Arai, J. Math. Anal. Apply. {\bf 158}, 63 (1991).

\bibitem{51} S.~Flügge, Practical Quantum Mechanics I, (Springer Verlag, Berlin,
1971).

\bibitem{52} A.~F.~Nikiforov, and V.~B.~Uvarov, Special Functions of
Mathematical Physics (Birkhauser, Basel, 1988).

\bibitem{53} C.~Tezcan, and R.~Sever, Int. J. Theor. Phys. {\bf 48}, 337 (2009) [arXiv:quant-ph/0807.2304].

\bibitem{54} M.~Abramowitz, and I.~Stegun. (Eds), Handbook of Mathematical Functions with Formulas,
Graphs and Mathematical Tables (New York, 1972).
\end{thebibliography}
\end{document}